\documentclass{emulateapj}
\usepackage{lscape} 
\usepackage{graphicx}
\usepackage{natbib}
\usepackage{rotating}
\newfont{\rten}{cmr10} 
\def \arcdeg{\hbox{$^\circ$}}

\begin{document}


\title{Could the planets around HR 8799 be brown dwarfs?}

\author{Amaya Moro-Mart\'in\altaffilmark{1,2}
George H. Rieke\altaffilmark{3}
Kate Y. L. Su\altaffilmark{3}}

\altaffiltext{1}{Departamento de Astrof\'isica, CAB (CSIC-INTA),  Instituto Nacional de T\'ecnica Aeroespacial, Torrej\'on de Ardoz, 28850, Madrid,  Spain}
\altaffiltext{2}{Department of Astrophysical Sciences, Princeton University, Peyton Hall, Ivy Lane, Princeton, NJ 08544, USA}
\altaffiltext{3}{Steward Observatory, University of Arizona,  933 North Cherry Ave, Tucson, AZ 85721, USA}

\begin{abstract}
We consider the limiting case for orbital stability of the companions to HR 8799. This case is only consistent with ages for the system of $\sim$ 100 Myr, not with the 1 Gyr age proposed from astroseismology. The discrepancy probably arises because the inclination of the star is smaller than assumed in analyzing the astroseismology data. Given this young age, the best estimates of the companion masses place them by a small margin on the planet side of the division between planets and brown dwarfs. 
\end{abstract}


\section{Introduction}
\label{intro}
HR 8799 is a late A/early F star located at 39.4 pc, with T$_{\rm eff}$ $\sim$ 7350K, M$_*$ = 1.5 M$_\odot$ and a $\lambda$ Bootis-like pattern of surface metallicity. The star has received 
a very high level of attention since the imaging of three massive substellar objects in
large radius orbits around it (Marois et al. 2008). HR 8799 and Fomalhaut (Kalas et al. 2008) 
are the only two firm examples of systems with separations $>$ 50 AU between the star and planetary-mass companions. Both stars also have luminous planetary debris disks that can provide further insight to the outer structures of their planetary systems (e.g., Stapelfeldt et al. 2004; Su et al. 2009). Given the bias in current planet-discovery techniques toward close companions,
the existence and properties of these widely separated examples open a new perspective on the diversity of planetary systems. 

The nature of the companions of HR 8799 depends critically on the age of the system, since their cooling rates (Baraffe et al. 2003) enter strongly into 
the relation of their luminosities to their masses. 
The star is near the zero-age main sequence, where traditional methods for estimating its age are not accurate. Marois et al. (2008) used a variety of arguments\footnote{The stellar Galactic space motion (UVW)  suggests that the star as a member of the Local Association at 20--150 Myr;  the excess emission at 24 $\mu$m from its debris disk also suggests a young system.} to estimate
an age of a $30 - 160$ Myr and on this basis assign masses  of $5 - 11$, $7 - 13$, and $7 - 13$ M$_{Jup}$ respectively to HR 8799 b, c, and d. 
Janson et al. (2010) used VLT/NACO to obtain spatially resolved spectroscopy of HR 8799 c from 3.88 to 4.10 $\mu$m. Although there are some discrepancies in the slope of the spectrum beyond 4 $\mu$m, it can generally be fitted
by a COND model from Baraffe et al. (2003) with T$_{\rm pl}$
= 1100 K, log g = 4.0 and R$_{pl}$ = 1.3 R$_{Jup}$. These characteristics
are consistent with the mass estimate for
planet c of 10 M$_{Jup}$ at 60 Myr by Marois et al. (2008). The minimum
mass for deuterium burning, $\sim$ 13 M$_{Jup}$, is often taken as the dividing 
line between planets and brown dwarfs, in which case these mass estimates
indicate that HR 8799 hosts a fascinating planetary system.

However, Moya et al. (2010a) argue that the stellar age is still unconstrained, pointing  out that some of the age estimators are compromised by the $\lambda$ Bootis character of the star or are statistical in nature. They used astroseismology to show that an older age of $\sim$ 1 Gyr is possible, if the inclination of the equatorial plane of the star is $I_* > 36 \degr$.  Because the masses of the companions are derived from the comparison of the observed luminosity to theoretical evolutionary tracks of giant planets and brown dwarfs, the age of the star is critical in determining these masses. At this older age, the companion luminosities would have faded significantly and their assigned masses would be well into the brown dwarf range. Thus, the nature of the HR 8799 system would change significantly. This would affect the formation scenarios under discussion, and the ability to use these three planets (of similar age and metallicity) to calibrate evolutionary and atmospheric models of giant planets. 

In this letter, we assess the possibility of more massive companions by considering the limiting case in which the masses of the planets are set to the largest values that dynamical stability can allow (Section \ref{planet_conf}). Given the observed planetary luminosities and using evolutionary models, we estimate the age corresponding to these limiting masses. We then evaluate the stability of the planets and the dust-producing planetesimals during that timescale to assess whether an older age could still lead to a planetary system (planets + planetesimals) consistent with the observations (Section \ref{stability}). We discuss the results in Section \ref{discussion}. 

\label{intro_planets}

\section{A stable configuration with massive bodies - a limiting case}
\label{planet_conf}

We now consider whether the configuration of the companions can help resolve the uncertainty over the age of HR 8799. The companions have been detected in HST/NICMOS archival data taken 10 years prior to the discovery image (Lafreniere et al. 2009),  however, their long orbital periods  make the determination of the orbital parameters still uncertain. While there are several configurations that would fit (1) the astrometry, (2) the observed luminosities, and (3) the requirement that the system is stable for at least 30-160 Myr, dynamical analysis by various authors converge on a solution in which the companions are on nearly circular orbits, locked into a double 4:2:1 mean motion resonance  (1d:2c:4b). This possibility was first suggested by Fabryck \& Murray-Clay (2010), and latter confirmed by Go{\'z}dziewski \& Migaszewski (2009) and Reidemeister et al. (2009). (For a summary of the possible orbital solutions see section 2.4.2 in Moro-Mart\'in et al. 2010). The companions avoid close encounters in the resonant configuration, ensuring dynamical stability. 

In this paper, we focus on this configuration because it may be the only solution that can guarantee the dynamical stability of the system on a timescale comparable to the stellar age, and in the case of massive planets.  The analysis in Fabrycky \& Murray-Clay (2010) concluded that, under this configuration, the planetary masses can be increased up to 1.9 times their nominal values (see their Figure 11). Therefore, we consider the following planetary masses with orbital elements listed in Table \ref{planetparam}: 1.27$\cdot$10$^{-2}$ M$_{\odot}$ (= 13.30 M$_{Jup}$) for planet b, and 1.80$\cdot$10$^{-2}$ M$_{\odot}$ (= 18.85 M$_{Jup}$) for planets c and d; this would put planets c and d well into the brown dwarf regime, while planet b would be near the transition. Following the labeling convection in Moro-Mart\'in et al. (2010), we will refer to this configuration as fit D6. 

\begin{deluxetable}{lccccccc}
\tablewidth{0pc}
\tablecaption{Planetary and stellar parameters for the dynamical simulations (Fit D6a)\label{planetparam}}
\tablehead{
\colhead{$M_{planet}$} &
\colhead{$a$} &
\colhead{$e$} &
\colhead{$i$} &
\colhead{$\Omega$} &
\colhead{$\omega$} &
\colhead{$M$}\\
\colhead{(M$_{\odot}$)} &
\colhead{(AU)} &
\colhead{} &
\colhead{(rad)} &
\colhead{(rad)} &
\colhead{(rad)} &
\colhead{(rad)}} 
\startdata
1.27$\cdot$10$^{-2}$ (b)	& 	67.91	&	0.002 		& 0  	&	0 	& $\pi$      & $\pi$ \\
1.80$\cdot$10$^{-2}$ (c)	& 	37.97  	&	0.005 		& 0  	&	0 	& 0 		& 0 \\
1.80$\cdot$10$^{-2}$ (d)	&	23.52  	&	0.083  		& 0  	& 	0  	& $\pi$      & 0 \\
\enddata
\tablecomments{Planetary parameters used in the dynamical simulation D6a (taken from Fabrycky \& Murray-Clay 2010). The planet masses correspond to 13.30, 18.84 and 18.84 M$_{Jup}$, for planets b, c and d, respectively. $\it{a}$ and $\it{e}$ are the semi-major axis and eccentricity of the planet; the orbits are face on  ($I_{\rm pl}$ = 0) and co-planar ($i$ = 0); $\omega$ is the argument of periastron,  $\Omega$ is the longitude of the ascending node, and $M$ is the mean anomaly.  The mass of the central stars is 1.5 M$_{\odot}$. 
}
\end{deluxetable}

The conclusion that the values adopted here are the maximum permitted masses is based on the analysis by Beauge et al. (2003): for a given pair of eccentricities ($e_1, e_2$),  they estimated the maximum mass of the innermost planet ($M_1$)  for which stable apsidal corotation at the 2:1 resonance can exist.  For the configuration considered in this paper (see Table \ref{planetparam}), where the planet pair b-c has $e_1 = e_c$ = 0.005 and $e_2 = e_b$ = 0.002, Beauge et al. (2003) estimated that the maximum mass of ${M_1 = M_c}$ is $\lesssim$ 0.012 M$_*$ (see their Figure 8); similarly, the planet pair c-d, with $e_1 = e_d$ = 0.083 and $e_2 = e_b$ = 0.005, would correspond to  ${M_1 = M_d}$ having a maximum mass of $\sim$ 0.013 M$_*$. These values are similar to those adopted in this paper for planets c and d (1.80 $\cdot$10$^{-2}$ M$_{\odot}$ = 0.012 M$_*$, for M$_*$= 1.5 M$_{\odot}$). Configurations with planetary masses more massive than these cannot have a stable corotation orbit in the 2:1 resonance. In other words, the fit D6 adopted in this study is a limiting case: the upper bound for the dynamically allowed planetary masses. 

Using different evolutionary models, and based on the luminosities reported by Marois et al. (2008),  log(${L}\over{L_{\odot}}$) = -5.1 for planet b and log(${L}\over{L_{\odot}}$) = -4.7 for planets c and d, we can estimate the age of the system that would correspond to the observed luminosities and the masses adopted in this paper: using the models of Chabrier et al. (2000), the age would be 100--500 Myr (it cannot be constrained any further because their tables do not include results for intermediate ages); from Baraffe et al. (2003 -- their Figure 1), we obtain an age of  250--350 Myr; while from Burrows et al. (1997 - their Figure 7) we get that if planet b were to have 1.27$\cdot$10$^{-2}$ M$_{\odot}$, its observed luminosity would correspond to a stellar age of $\sim$ 260 Myr. The temperature of 1100 K for HR 8799 c (Janson et al. 2010) can also be used to constrain the age: from Chabrier et al. (2000 - their Figure 2), this temperature and adopted mass would correspond to an age of $\lesssim$ 225 Myr, while from Baraffe et al. (2003 - their Figure 1) the age would be in the range 250--280 Myr. We conclude that if the planets were to have the limiting masses adopted in this paper (1.9 times their nominal values), their current luminosities would imply a system age of approximately 250--350 Myr. 

\section{Stability of companion and dust-producing planetesimal orbits}
\label{stability}

We now evaluate the stability of the planets and the dust-producing planetesimals on the timescales estimated above. Following Fabrycky \& Murray-Clay (2010), and to assess how sensitive the results are to small changes in initial conditions,  we consider several planetary configurations around the one described in Section \ref{planet_conf} and Table \ref{planetparam}; the differences are 10$^{-4}$ AU for the semi-major axes, 10$^{-4}$ for the eccentricities, and 0.01$\degr$ for the inclinations, ascending nodes, longitudes of periastron, and mean anomalies. In total, we model 14 different sets of initial conditions for the orbits.  In addition to the three companions and the central mass of 1.5 M$_{\odot}$, the simulations include 500 test particles. Seven of the simulations consider particles in the inner disk,  uniformly distributed between 2 AU and the semi-major axis of the inner-most planet, while the remaining seven consider particles in the outer disk,  spaced uniformly between the semi-major axis of the outermost planet and 300 AU. We assume that the planets and the dust-producing planetesimals formed out of a thin disk and are co-planar.  The particles are on initially circular orbits, with angular elements chosen randomly between 0 and 2$\pi$. Particles are removed if they approach the star closer than 2 AU, or approach a planet closer than its Hill radius. The orbits were integrated using the multiple time step symplectic method skeel-SyMBA (Duncan, Levison \& Lee 1998).

\begin{figure}
\begin{center}
\includegraphics[scale=0.55,angle=0]{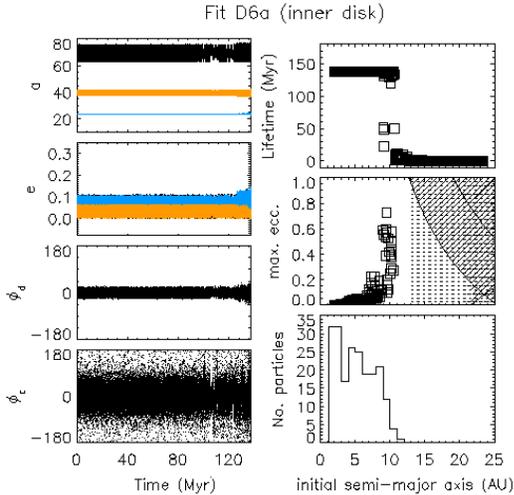}
\end{center}
\caption{{\it Left}: Long-term evolution of the planets' orbits under simulation D6a, during the time in which the system is stable. The orbital elements are listed in Table \ref{planetparam}. From top to bottom, the panels correspond to semi-major axis ($a$), eccentricity ($e$) and resonant angles for the 2:1 resonance between planets c and d ($\phi_d$), and the 2:1 resonance between planets b and c ($\phi_c$).  The mean and standard deviation of the resonant angles are listed in Table \ref{lifetime}. The colors correspond to planet b (black), c (orange) and d (blue). {\it Right}: Results from the dynamical simulation of 500 test particles in the inner planetesimal disk.  Top: test particle's lifetimes. Middle: allowed parameter space for the planetesimals' orbital elements, where the shaded areas indicate regions where test particle's orbits are unstable due to planet-crossing ({\it striped area}) or overlapping first order mean motion resonances ({\it dotted area}); the squared symbols show the maximum eccentricity attained by test particles on initially circular orbits.  Bottom: number of surviving test particles at the onset of dynamical instability.} 
\label{ploe1}
\end{figure}

\begin{figure}
\begin{center}
\includegraphics[scale=0.55,angle=0]{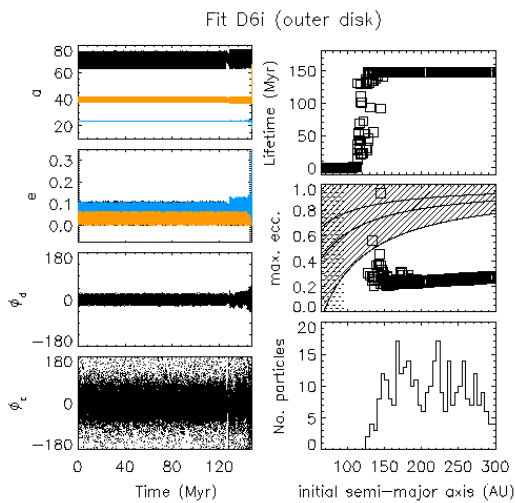}
\end{center}
\caption{
 Same as Figure \ref{ploe1} but for a simulation of 500 test particles in the outer planetesimal disk. In this case, the high baseline value of the test particles' eccentricities and slight rising trend is a numerical artifact from using stellar-centric osculating elements.} 
\label{ploe2}
\end{figure}

\subsection{Planets}
\label{stability_planets}
 
Figures \ref{ploe1} and \ref{ploe2} show the evolution of the planets' semi-major axes, eccentricities, and resonant angles for the 2:1 resonance between planets c and d ($\phi_d$),  and the 2:1 resonance between planets b and c ($\phi_c$), for the two longest-lived simulations, one for the inner disk and one for the outer disk, with dynamical lifetimes of 137.5 Myr and 147.5 Myr, respectively. 
While the companions are locked into the double resonance, the system is stable. 
Table \ref{lifetime} lists the resonant angles (mean and standard deviation) and lifetimes of the 14 simulations ran, the latter ranging from of 9 to 147 Myr. 
Using the MERCURY integrator, Fabrycky \& Murray-Clay (2010) found that the time to instability of their six simulations (with the same set of initial conditions) range from 25.9 Myr to 102.6 Myr.
We conclude that if the planets were to have the limiting masses adopted in this paper (1.9 times their nominal values), the lifetime of the system would be $\lesssim$ 150 Myr.  

\subsection{Dust-producing planetesimals}
\label{stability_planetesimals}

Following Moro-Mart\'in et al. (2007, 2010), we have estimated the potential location of the dust-producing planetesimal belts under the configurations considered in this paper. The possible niches are the regions where the orbits are stable and the maximum eccentricities excited by the planetary perturbations remain low ($e_{\rm max} \lesssim$ 0.3), ensuring the survival of planetesimals against quick erosion from mutual collisions.  Figures \ref{ploe1} and \ref{ploe2}  show the results for the two longest-lived simulations already discussed (one for the inner disk, and one for the outer disk). We find that the outer edge of the inner (warm) planetesimal disk is R$_{\rm out}^{\rm warm} \sim $ 9 AU, and that the inner edge of the outer (cold) planetesimal disk is R$_{\rm in}^{\rm cold} \sim$ 150 AU. We have adjusted the disk model presented in Su et al. (2009) to this new geometry and find that a satisfactory fit to the spectral energy distribution can be achieved readily. (For comparison, the configuration adopting the nominal masses of 7, 10 and 10 M$_{Jup}$ for planets b, c and d, respectively -- labeled  D1 in  Moro-Mart\'in et al. 2010 -- results in R$_{\rm out}^{\rm warm} \sim $ 12 AU and R$_{\rm in}^{\rm cold} \sim $ 110 AU). We conclude that the presence of planets with masses 1.9 times their nominal values would still be consistent with the  SED and imagining of the debris disk around HR 8799. 

\begin{deluxetable*}{lcccccccccccccc}
\tablewidth{0pc}
\tablecaption{Results from dynamical simulations\label{lifetime}}
\tablehead{
\colhead{Model} &
\colhead{D6a} &
\colhead{D6b} &
\colhead{D6c} &
\colhead{D6d} &
\colhead{D6e} &
\colhead{D6f} &
\colhead{D6g} &
\colhead{D6h} &
\colhead{D6i} &
\colhead{D6j} &
\colhead{D6k} &
\colhead{D6l} &
\colhead{D6m} &
\colhead{D6n}}
\startdata
Dyn. Lifetime (Myr)						 				& 	137.5 	& 	41.2 	& 	110.9 & 	59.3 	& 	29.9 	& 86.5 	& 	32.0 	& 	117.3 & 	147.5 & 	43.3 	& 	19.2 	& 	9.1 		& 	17.6 & 	31.8 \\
mean(${\phi_c}$) ($\degr$)								&	-0.04		& 	-0.84	&	0.14	&	-0.50&	6.52	& -0.06	&	-4.82&	0.86	&	0.19	&	-1.64	&	1.03	& 	0.29		& 	0.67	&	0.56 \\
$\sigma(\phi_c$) ($\degr$)								& 	44.4		&	45.7	&	44.6	&	43.9	&	48.7	& 44.1	&	48.9	&	44.3	&	43.4	&	46.4	&	103.2  & 	103.5	& 	103.1 &	102.5\\
mean($\phi_d$) ($\degr$)									&	-0.08		&	-0.17	&	-0.20	&	-0.08	&	1.90	& 0.03	&	-0.96	&	0.24	&	0.01	&	-0.13	&	-5.67	     & 	3.47		& 	0.80	&	-0.56\\
$\sigma(\phi_d)$ ($\degr$)								& 	8.9		&	9.9	&	9.2	&	9.4	&	10.3	& 9.0		&	9.9	&	9.2	&	8.6	&	10.1	&	103.0 & 	104.2  	& 	101.0 &	102.0\\
\enddata
\tablecomments{Results for the dynamical simulations around the configuration described in Table \ref{planetparam}.  
${\it Dyn.~Lifetime}$ is the time for instability in Myr. 
${\it Mean({\phi_c})}$ and ${\it mean({\phi_d})}$ are the mean of the resonant angles for the 2:1 resonance between planets b and c ($\phi_c$ ),  and the 2:1 between planets c and d ($\phi_d$), during the time in which the system is stable. $\phi_c = 2\lambda_b-\lambda_c-\tilde{\omega}_c$ and $\phi_d = 2\lambda_c-\lambda_d-\tilde{\omega}_d$, where $\lambda$ is the mean longitude, $\lambda = M + \Omega + \omega$, and $\tilde{\omega}$ is the longitude of pericenter, $\tilde{\omega} = \Omega + \omega$. $\sigma(\phi_c$) and $\sigma(\phi_d)$ are the standard deviations of the resonant angles. The test particles in models D6a--e, k and m are initially located in the inner disk, while the test particles in models D6f--j, l and n are located in the outer disk.}
\end{deluxetable*}

\section{Discussion and Conclusion}
\label{discussion}

The configuration considered in this paper (D6) is a limiting case, in the sense that the adopted companion masses are the maximum values for which the stabilization mechanism, the double 4:2:1 resonance, can apply. Based on stability considerations, it is therefore unlikely that the companions are more massive than considered here. 

In Moro-Mart\'in et al. (2010) we found that configurations in which the planets adopt their nominal masses (7, 10 and 10 M$_{Jup}$ for planets b, c and d, respectively -- labeled D1, D4 and D5) are consistent with planet and dust observations because: (1) the planets masses correspond to an age estimate of $\sim$ 60 Myr and the system is found to be stable for $>$ 160 Myr; and (2) the orbits of the dust-producing planetesimals fulfill two requirements: (2a) they are in agreement with the dust location as derived from debris disk observations; and (2b) for the timescale during which the planet orbits are stable, the planetesimal orbits are also  stable and their eccentricities remain low (implying survival against erosion). In contrast, the configuration considered in this paper, where the planets are 1.9 times more massive, is not fully consistent.  Conditions (2a) and (2b) above are fulfilled: for the timescale during which the planets are on stable orbits,  there are planetesimals in the regions where the dust is inferred to be located, and the planetesimal orbits are stable and their eccentricities remain low. However, condition (1) is violated:  while the substellar object evolutionary models predict an age of the system around 250--350 Myr (Baraffe et al. 2003) or $\sim$ 260 Myr (Burrows et al. 1997) from luminosity constraints, or  $\lesssim$ 225 Myr (Chabrier et al. 2000) and 250--280 Myr (Baraffe et al. 2003) from temperature constrains, the dynamical lifetime of the system is significantly smaller,  $\lesssim$ 150 Myr.  Based on dynamical considerations we therefore conclude that the large companion masses considered in this paper are unlikely and that the dynamical state of the system favors ages of $\sim$ 100 Myr and companion masses slightly below the brown dwarf regime.  

This conclusion is supported by the strong emission of the debris disk at 24 $\mu$m, behavior that is very uncommon around stars older than $\sim$ 500 Myr (Gaspar et al. 2009). In addition, the average temperature of 7350 K and the luminosity of 5 L$_\odot$ from Sadakane (2006) and Moya et al. (2010b) places the star at or below the zero age main sequence and at a very young age compared with other $\lambda$ Boo stars (Paunzen et al. 2002). 

However, these results are not consistent with the possibility that the system is $\sim$ 1 Gyr old. The astroseismology modeling that suggests this age (Moya et al. 2010a) depends critically on the rotation speed of the star and hence on the assigned inclination. It is only valid if the inclination of the equatorial plane of the star is $I_* \ge $36$\degr$. The case for $I_* = 36 \degr$ results in two sets of possible solutions, a stellar mass of 1.32--1.33 M$_{\odot}$ and an age of 1123--1625 Myr, or a stellar mass of 1.44--1.45 M$_{\odot}$ and and age of 26--430 Myr (with 16.7\% of the models lying in the generally adopted age range of 30--160 Myr). For $I_* = 50\degr$, Moya et al. (2010a) obtain a stellar mass of 1.32 M$_{\odot}$ and and age of 1126--1486 Myr. 

However, Lafreniere et al. (2009) suggested that the inclination of the orbital plane of the planets is $I_{\rm pl}$ = 13--23$\degr$ from astrometry observations with a 10 year baseline (using NICMOS archival data). Su et al. (2009) reported from the degree of symmetry of the spatially resolved 70 $\mu$m disk that the inclination of the disk is consistent with being face-on and unlikely to be $I_{\rm disk} >$ 25$\degr$. We have re-evaluated these observations and place a 3$\sigma$ upper limit to $I_{\rm disk}$ of 40$\degr$. Upcoming JCMT/SCUBA-2, Herschel and HST imaging of HR 8799 will be able to further constrain the inclination of the debris disk. 
If we assume that the equatorial plane of the star coincides with the orbital plane of the planets and the dust-producing planetesimals (i.e. $I_{*} = I_{\rm pl} = I_{\rm disk}$), these observational constraints indicate that indeed the inclination of the star is likely inside the 18 $\degr < I_* < 36 \degr$ range where the astroseismology models cannot be applied, or that it is close to 36$\degr$ where they permit solutions consistent with the $\sim$ 100 Myr age derived from other considerations. 

We conclude that HR 8799 is young, with an age of $\sim$ 100 Myr. Its companions are therefore likely all to be very massive planets, just below the mass limit for deuterium burning. 

\begin{center} {\it Acknowledgments} \end{center}
We thank Hal Levison for providing  skeel-SyMBA  for the dynamical simulations and M. Janson, D. Fabrycky, R. Murray-Clay, A. Moya and P. Kalas  for useful discussion. A.M.M. acknowledges funding from the Spanish MICINN (Ram\'on y Cajal Program RYC-2007-00612, and grants AYA2009-07304 and Consolider Ingenio 2010CSD2009-00038).

\end{document}